# Single-Phase *L*1$_0$-Ordered High Entropy Thin Films with High Magnetic Anisotropy

*Willie B. Beeson, Dinesh Bista, Huairuo Zhang, Sergiy Krylyuk, Albert V. Davydov, Gen Yin, and Kai Liu* [*]

W. B. Beeson, D. Bista, G. Yin, K. Liu

Physics Department, Georgetown University, Washington, D.C. 20057, USA

E-Mail: kai.liu@georgetown.edu

H. Zhang, S. Krylyuk, A. Davydov

Materials Science and Engineering Division, National Institute of Standards and Technology, Gaithersburg, MD 20899, USA

H. Zhang

Theiss Research, Inc., La Jolla, CA 92037, USA

**Abstract**

The vast high entropy alloy (HEA) composition space is promising for discovery of new material phases with unique properties. This study explores the potential to achieve rare-earth-free high magnetic anisotropy materials in single-phase HEA thin films. Thin films of FeCoNiMnCu sputtered on thermally oxidized Si/SiO$_2$ substrates at room temperature are magnetically soft, with a coercivity on the order of 10 Oe. After post-deposition rapid thermal annealing (RTA), the films exhibit a single face-centered-cubic phase, with an almost 40-fold increase in coercivity. Inclusion of 50 at.% Pt in the film leads to ordering of a single *L*1$_0$ high entropy intermetallic phase after RTA, along with high magnetic anisotropy and 3 orders of magnitude coercivity increase. These results demonstrate a promising HEA approach to achieve high magnetic anisotropy materials using RTA.

# 1. Introduction

High entropy alloys (HEAs) are a class of materials traditionally defined to contain 5 or more elements which may exist as stable or metastable single phases due to their high configurational entropy.[1] The entropy-stabilization effect implies a vast number of unexplored material systems with potential to exhibit unconventional properties. So far HEA studies have predominantly focused on exceptional mechanical properties, namely combinations of strength, hardness, and ductility.[1-2] In recent years, however, a number of systems have been reported which exhibit other attractive properties and functionalities, which are suitable for hydrogen storage,[3] thermoelectrics,[4] superconductivity,[5] magnetocalorics,[6] and soft magnet applications.[7] Magnetic studies make up comparatively little of the published literature, and most HEA studies have been limited to bulk systems.

A prospect which remains largely unexplored is the design of HEAs with high magnetic anisotropy, which have critical applications in magnetic recording and permanent magnets.[8] As the search for new rare-earth-free high anisotropy materials has mostly exhausted the pool of binary and ternary alloy compositions, an emerging opportunity is the arena of high entropy materials, which may hold unique electronic structures favorable to high magnetic anisotropy. Most known HEAs fall in the category of soft magnetic materials, with coercivities below 100 Oe.[9] However, coercivities as high as 1200 Oe have been reported in certain HEA systems.[10] The coercivity is highly influenced by microstructure which may differ significantly amongst HEA systems depending on the fabrication conditions. The bulk of HEA studies focus predominantly on the maximum entropy solid solutions at the compositional center, which generally consist of uniform chemical disorder within cubic crystal structures, while low-symmetry crystal structure and chemical order play an important role in magnetocrystalline anisotropy. However, it is not necessary that HEAs be fully disordered, as the configurational entropy can be a dominant term in the free energy as long as the number of elements is large relative to the number of unique lattice sites. Certain HEA compositions may exhibit long-range chemical order through the emergence of multiple sublattices, analogous to high entropy ceramics,[11] leading to so-called "high entropy intermetallic" (HEI) compounds.[12] Previously, transition metal alloys with high anisotropy, such as the $L1_0$ phases of FePt and FePd have been achieved by annealing of equiatomic solid solution pre-cursors.[13] In the case of metastable high anisotropy phases such as $L1_0$ FeNi, unconventional annealing techniques have been employed.[14] In particular, rapid thermal annealing (RTA) of transition metal alloy films on Si/SiO$_2$ is an effective means of inducing high anisotropy phase formation in solid solution precursors, in part due to the substrate-induced strain. [15]

In this study, we experimentally demonstrate two thin film HEAs with large coercivity and magnetic anisotropy achieved by RTA. We show that RTA treatment of equiatomic FeCoNiMnCu thin films leads to ~ 40-fold increase in coercivity. By introduction of 50 at.% Pt, we realize an *L*1$_0$ HEI phase of (FeCoNiMnCu)Pt with perpendicular magnetic anisotropy resulting from a reduced lattice symmetry and strong spin-orbit coupling. A significant increase of coercivity, by 3 orders of magnitude, is observed.

## 2. Results and Discussion

FeCoNiMnCu films were sputtered from a composite target (Experimental Methods), and a composition of Fe$_{0.19}$Co$_{0.18}$Ni$_{0.20}$Mn$_{0.18}$Cu$_{0.25}$ was confirmed by energy dispersive X-ray (EDX) microanalysis. Grazing incidence XRD (GIXRD) patterns of the 50 nm FeCoNiMnCu films sputtered at various substrate temperature (T$_s$) are shown in **Figure 1a**, along with those sputtered at 20 °C and treated with RTA for 30 s at 600 °C. The films sputtered at 20 °C exhibit only one peak at ~44° with relatively low intensity. A similar pattern is seen in the films grown at 350 °C (**Figure S2**). For the film grown at T$_s$ = 500 °C, new peaks are observed, corresponding to a face-centered-cubic (*fcc*) phase with lattice parameter of 3.58 Å and a body-centered-cubic (*bcc*) phase with lattice parameter of 2.84 Å. With further increase of T$_s$, the peak intensities increase overall and the *fcc* (111) and *bcc* (110) peaks become more separated in 2$\theta$, revealing the increasing chemical segregation between the two phases. At T$_s$ = 700 °C, the lattice constants for the *fcc* and *bcc* phases reach 3.63 Å and 2.86 Å, respectively, close to the bulk values of *fcc* Cu (3.615 Å) and *bcc* Fe (2.867 Å). Considering that Fe-Cu has a binary mixing enthalpy ΔH$_{mix}$ of 12.9 kJ/mol, which is the most positive of all binary pairs in the FeCoNiMnCu system (**Table S1**), we expect it to be a dominant driving force of phase separation.[16] Based on this, we consider the separated phases most likely to be a Cu-rich *fcc* phase and an Fe-rich *bcc* phase. In the films grown at 500 °C, 600 °C, and 700 °C, there are additional peaks just below 40° and below 60° which originate from Ta capping layer used for their growth.

In contrast, the films sputtered at 20 °C and 350 °C with post-deposition RTA at 600 °C for 30 s exhibit a single-phase *fcc* pattern (**Figure S1** and **S2**). This is in agreement with previous reports of bulk FeCoNiMnCu high entropy alloys [6c, 17] as well as the phase formation criteria based on the valence electron concentration (VEC).[18] The formation of single-phase *fcc* in the RTA film, as opposed to multi-phase *fcc* + *bcc*, demonstrates the advantage of the RTA method in suppressing growth of secondary phases, since the magnitude of the entropic term in the free energy decreases with temperature, leading to secondary phase growth at intermediate temperatures.[19] The 13 nm thick films sputtered at 350 °C were treated with RTA at 600 °C for additional dwell times. Those annealed for a shorter duration of 10 s and 15 s exhibit *fcc* peaks in addition to a minor secondary phase peak near 45° (Figure S2), which is consistent with the (110) peak of the *bcc* phase seen in the films grown at high temperature. After 20 s of RTA, only the *fcc* phase is detected.

The film surface morphology was probed using scanning electron microscopy (SEM) and atomic force microscopy (AFM). The upper panels of Figure 1b show the SEM images of film surface for increasing $T_s$, where a large change in morphology is observed. The films grown at $T_s$ up to 350 °C are flat and continuous (not shown). At $T_s$ = 500 °C, the film roughness increases, exhibiting a granular structure of particles with an average diameter of ~20 nm. For $T_s$ of 600 °C – 700 °C, the films break up into islands, likely due to dewetting, with average size increasing from ~50 nm to 80 nm. A similar effect occurs for prolonged RTA times at 600 °C, as illustrated in the SEM images in the lower panels of Figure 1b for 13 nm thick films grown at 350 °C. The film remains smooth and continuous up to 15 s of RTA. After 20 s of RTA, areas of dark contrast emerge, corresponding to void formation in the film which is confirmed by AFM (**Figure S3**). After RTA for 30 s, the voids grow and connect to a 45% area fraction. Similar void formation has been reported before during RTA of thin metallic films, attributed to strain relaxation, change in unit cell volume upon phase transformation, and changes in surface energy.[15c, 15d, 20] In particular, the high heating rate in RTA has been shown to generate large residual stress in thin films which is relieved upon dewetting.[15d] For the 50 nm thick films treated with RTA at 600 °C for 30 s (Figure S3), the voids have comparable number density but much smaller size, making up <5% area fraction, indicating the suppression of void growth kinetics in the thicker film.

Sample magnetic properties were measured by vibrating sample magnetometry (VSM) and superconducting quantum interference device (SQUID) magnetometry (**Table S2**). The films grown at 350 °C exhibit a saturation magnetization $M_s$ = 595±45 emu cm$^{-3}$ ($\sigma_s$ = 70±6 emu/g) as-grown and 590±30 emu cm$^{-3}$ after RTA for 30 s, lying at the upper end of literature values of 20-80 emu g$^{-1}$ for bulk FeCoNiMnCu systems.[6c, 21] The wide range of reported $M_s$ suggests that the stability of long-range ferromagnetic (FM) order and a large net magnetic moment can be sensitive to small changes of the film composition and the lattice constant.[22] Particularly, Mn is expected to reduce the magnetization due to antiferromagnetic (AF) nearest-neighbor exchange interactions, whereas addition of Cu to FeCoNiMn is suggested to stabilize a FM order.[21b] To gauge the sensitivity of $M_s$ on the presence of non-FM Mn and Cu in the present system, we have performed density functional theory (DFT) calculations of the magnetic moment per Mn atom in $Fe_{0.19}Co_{0.18}Ni_{0.20}Mn_xCu_{0.43-x}$ for different Mn:Cu ratios (**Figure S4**). The calculations used the thin-film lattice constant obtained by XRD, and the compositions used in the calculations are chosen based on the EDX measurement from our samples. The DFT results reveal a FM ground-state ordering of Mn-Mn for $x \leq 0.25$ and AF ordering for $x > 0.25$, suggesting that the net $M_s$ can vary significantly even near the equiatomic composition in our thin film. Also, the maximum magnetic moment of Mn is found close to the experimental Mn:Cu ratio in our films. This suggests that the sizable $M_s$ in our system may be partly attributed to the lower Mn:Cu ratio which promotes FM ordering and high moment of Mn. Further increasing the substrate temperature ($T_s > 500$ °C) leads to an increased $M_s$ likely due to separation of the Fe-rich phase observed in XRD (Fig. 1a), reaching a value of $675\pm25$ emu cm$^{-3}$ at $T_s = 700$ °C.

To better illustrate the evolution of sample magnetic characteristics with changing RTA conditions, we have performed first-order reversal curve (FORC) analysis in the in-plane geometry,[23] as shown in **Figure 2**. The as-grown films sputtered at 350 °C are magnetically soft, with a small coercivity of 10 Oe (Figure 2a), and the FORC distribution exhibits a single peak near local coercivity $H_c = 0$, revealing mostly reversible switching (Figure 2e).[24] After RTA at 600 °C for 10 s, the sample develops a sizable hysteresis, with a coercivity of 170 Oe (Figure 2b). The FORC distribution shows a separate peak centered around $H_c = 150$ Oe, indicating irreversible switching in the film (Figure 2f). Additionally, the previous peak near $H_c = 0$ still persists, but becomes spread along the $H_b$ axis, which indicates reversible magnetization switching due to thermal or anisotropic demagnetization effects.[24-25] After 15 s and 30 s of RTA, the samples exhibit an even wider hysteresis with a coercivity of 350 Oe and 390 Oe (Figure 2c and 2d), respectively. The primary FORC peaks centered at $H_c = 400$ Oe and 450 Oe, respectively, are clearly distinct from the reversible ridge near $H_c = 0$, while the latter feature becomes more spread along the $H_b$ axis (Figure 2g and 2h). By numerical integration of the FORC distribution around the irreversible switching feature, we find that it is associated with 60% of the magnetization in Figure 2h.

The 1-2 orders of magnitude coercivity increase in RTA-treated samples, compared to the as-grown ones, suggests the emergence of sizeable magnetic anisotropy in the film. The fact that the continuous 15s RTA sample (Figure 1b) exhibits a 350 Oe coercivity, indicates that the dewetting process and film morphology evolution in some of the samples are not essential to the significant coercivity enhancement. Previous studies on *bulk* FeCoNiMnCu solid solutions have not found appreciable magnetic anisotropy ascribed to the disordered phase.[21a, 21c, 26] Hard magnetic properties in some solid solutions have been attributed to the magnetocrystalline or shape anisotropy of a secondary phase, however studies of bulk phase-separated FeCoNiMnCu have not shown large coercivity of the order presented here.[21a, 21c] In the present RTA-treated samples, likely the anisotropic strain has induced lattice and grain distortions or directional chemical order, different from those in bulk systems, and stabilized sizable magnetic anisotropy, similar to that in RTA-assisted synthesis of $L1_0$ FeCuPt.[15c, 27]

To gain further insight into the coercivity mechanism, angular dependence of the coercivity as a function of RTA time has been measured and well-fit to a model based on domain wall displacement (**Figure S5**), indicating a pinning-controlled coercivity for the in-plane direction.[28] The evolution of fitting parameters also reveals significant deviations in the overall demagnetization factors from the thin film geometry after RTA, even before film dewetting, reflecting the change in anisotropy induced by RTA. Therefore, the hard magnetic properties in FeCoNiMnCu film may be explained by RTA-induced microstructure changes and anisotropy enhancement which strengthens domain wall pinning.

While the combination of saturation magnetization and coercivity for these RTA- treated FeCoNiMnCu films is impressive in comparison to previously reported bulk systems, the coercivity is still modest, due to the high symmetry of the cubic lattice which limits the intrinsic magnetocrystalline anisotropy. The facilitation of low-symmetry chemical order and crystal structure combined with strong spin-orbit coupling is in general necessary to realize strong magnetocrysalline anisotropy. To achieve these conditions in a high entropy system, we have added 50 at.% Pt to the FeCoNiMnCu alloy, employing the high entropy intermetallic design approach in search of highly anisotropic $L1_0$ ordering. Although the certainty of Pt sites reduces the configurational entropy compared to the equiatomic case, we demonstrate below that such optimized entropy can still stabilize a single high-entropy $L1_0$ phase.

Thin films of 20 nm thick (FeCoNiMnCu)Pt were sputtered as described in the Experimental Methods section. EDX analysis revealed an average film composition of $Fe_{0.11}Co_{0.12}Ni_{0.10}Mn_{0.09}Cu_{0.10}Pt_{0.48}$. The as-grown films are in a disordered *fcc* (A1) phase with a lattice parameter of 3.80 Å (not shown). After RTA for 60 s at 600 °C, $L1_0$ ordering appears in GIXRD and $\theta$-$2\theta$ (1° $\omega$-offset) scans (**Figure 3a**), evidenced by the emergence of the forbidden (001) superlattice peak and peak-splitting due to the tetragonal symmetry. The extracted lattice parameters of the $L1_0$ phase are $c = 3.60$ Å and $a = 3.83$ Å, respectively. The presence of $L1_0$ peaks in the GIXRD scan shows the existence of randomly oriented $L1_0$ grains. However, the $\theta$-$2\theta$ XRD reveal dominant (00*l*) reflections, with a narrow (001) rocking-curve (full-width-at-half-maximum of 2°), indicating a preferred (001) orientation. The $L1_0$ order parameter can be extracted from the ratio of integrated intensities of the superlattice and fundamental peaks according to $S = \sqrt{\frac{\left(I_{001}/I_{002}\right)_{exp.}}{\left(I_{001}/I_{002}\right)_{calc.}}}$ where the numerator includes the (001) and (002) intensities extracted from the $\theta$-$2\theta$ XRD data, and the denominator includes the calculated intensities.[29] A high *S*-value of 0.8 is found, indicating the presence of significant $L1_0$ ordering, illustrated in Figure 3b.

High angle annular dark field scanning transmission electron microscopy (HAADF-STEM) and selected area electron diffraction (SAED) were further carried out to characterize the microstructures of the (FeCoNiMnCu)Pt films. As shown in Figure 3c, HAADF-STEM imaging on the as-grown (FeCoNiMnCu)Pt film shows a uniform grain size around 5 nm – 10 nm. SAED analysis (Figure 3d) shows that the as-grown film was crystallized into a well-defined *fcc* structure. After RTA processing, the grains grew substantially from several tens of nm up to 300 nm and transformed into $L1_0$ structure, indicated in SAED by the appearance of additional reflections forbidden for the *fcc* phase, such as the (001), (110), (201), (112) and (310) reflections (Figure 3e-3f). The large grain growth relative to some RTA-treated FePt films may be related to the inclusion of Cu which is known to promote diffusion and reduce the kinetic ordering temperature.[15a, 30] Atomic resolution HAADF-STEM images taken along the [001] and [110] zone-axis not only reveal the 4-fold and 2-fold symmetry, respectively, but also illustrate the striking chemical ordering of the $L1_0$ phase, apparent by the Z-contrast between FeCoNiMnCu and Pt sites (Figure 3g-3h). Further EDX analysis was conducted on approximately 50 grains and found no significant composition variations between grains or existence of binary phases. Therefore, we consider the film to be a single-phase high entropy intermetallic, with the 3$d$ transition metals ordering onto a high entropy sublattice of the $L1_0$ structure, opposite to Pt.

Magnetic properties of the (FeCoNiMnCu)Pt films have been studied by magnetometry and FORC. The as-grown films are magnetically soft with in-plane anisotropy and coercivity of 5 Oe (not shown). After RTA for 60 s at 600 °C, strong magnetic anisotropy has emerged, as shown by the families of FORCs and the corresponding FORC distributions in **Figure 4**. Large coercivity increases to 2.50 kOe and 1.46 kOe are observed for the in-plane (IP) and out-of-plane (OOP) loop (Figure 4a inset), respectively, which are also delineated by the outer boundaries of the FORCs (Figure 4a-4b). The high $M_r/M_s$ in the OOP loop relative to IP indicates the perpendicular anisotropy resulting from the (001) texture, as shown by the inset to Figure 4a. For the IP geometry, significant irreversible switching is revealed by the FORC peak centered at $H_c$ = 2.2 kOe, arising from high anisotropy $L1_0$ grains with IP easy-axis component (Figure 4c). Integrating this feature shows that it corresponds to 45% of the magnetization, while the remaining magnetization exists in the reversible ridge, representing magnetization reversal along the hard-axis of (001)-textured grains. For the OOP geometry, the FORC distribution exhibits two main features: a sharp peak centered at $H_c$ = 0.2 kOe and a broader horizontal ridge along the $H_c$ axis centered at $H_c$ = 2.0 kOe (Figure 4d). The FORC feature near the origin is notably different from a typical low anisotropy phase, which may be manifested as a vertical ridge along the $H_b$ axis due to shape anisotropy, as seen in the OOP FORC distribution of FeCuPt thin films containing A1 phase.[24b] Here, the low field FORC feature corresponds to a square loop contribution in the OOP direction (Figure 4c) despite the shape anisotropy energy, suggesting that it stems from a phase of sizeable perpendicular magnetic anisotropy. This feature likely originates from the large $L1_0$ grains seen in TEM (Figure 3e) which exhibit lower coercivity due to the absence of strong pinning centers such as grain boundaries and higher probability of surface nucleation sites. For example, bulk single-crystal $L1_0$ FePt may exhibit negligible coercivity despite its enormous anisotropy, due to the ease of domain wall motion through bulk single-crystals.[31] Meanwhile, grains with smaller sizes or varying $c$-axis orientation promote hysteresis through pinning at their grain boundaries. Integration of the OOP FORC feature at higher $H_c$ yields 45% of the saturation magnetization, equivalent to the primary FORC feature in the IP geometry. We propose that these high $H_c$ FORC features are associated with small and randomly oriented grains, which are evident by the observation of $L1_0$ peaks in GIXRD (Figure 3a). Therefore, the bimodal coercivity distribution observed in the OOP direction is explained by a 2-step process of magnetization reversal beginning with easy domain nucleation and propagation in the large (001) oriented grains (narrow low $H_c$ peak), followed by pinning between grains with small size and varying easy-axis direction (broad high $H_c$ peak). The distribution of size and easy-axis orientation both manifest in the spreading of the high $H_c$ peak along the $H_c$ axis. Based on this, we expect that the $H_c$ could be enhanced through microstructural optimization such as a reduction in grain size, which may be achieved by adjusting the RTA time and temperature.

An effective uniaxial anisotropy constant of $K_u = 2\times10^6$ erg cm$^{-3}$ was extracted from the area between IP and OOP hysteresis loops after correction of a shape component $2\pi M_s^2$. While not representative of the intrinsic magnetocrystalline anisotropy due to the polycrystalline microstructure, we note that $K_u$ extracted here is already in the range of moderate high magnetic anisotropy materials, although still somewhat lower than the intrinsic anisotropy of the binary $L1_0$ FePt and CoPt alloys, which are on the order of $10^7$ erg cm$^{-3}$. The saturation magnetization $M_s$ of 240±25 emu cm$^{-3}$ for the Pt-inserted film, which is roughly half of the $M_s$ obtained for the FeCoNiMnCu films, is reduced in comparison to the bulk values of FePt and CoPt. The $K_u$ of Pt-based $L1_0$ phases are determined to first-order by the effective VEC of the 3$d$ metal sublattice (i.e., Fermi level).[32] Based on the experimental composition, the reported $L1_0$ HEA films have an effective VEC of 9.4, and their $K_u$ is reasonably consistent with that of $L1_0$-ordered (Fe$_{0.25}$Ni$_{0.75}$)Pt and (Co$_{0.5}$Ni$_{0.5}$)Pt films having VEC of 9.5, which exhibit $K_u$ on the order of $10^6$ erg cm$^{-3}$.[33]

## 3. Conclusion

In summary, we have achieved high magnetic anisotropy in rare-earth-free high entropy metallic thin films by sputtering and RTA. Robust magnetic order of both *fcc* FeCoNiMnCu and $L1_0$ (FeCoNiMnCu)Pt films has been demonstrated. The FeCoNiMnCu films exhibit a large coercivity increase after RTA, resulting from microstructural changes and strain-induced anisotropy enhancement which strengthens domain wall pinning. Addition of Pt reduces the symmetry of the cubic *fcc* lattice to an intermetallic tetragonal $L1_0$ phase of (FeCoNiMnCu)Pt. Although the Pt insertion lowers the configurational entropy compared to the equiatomic case, the entropy is still high enough to stabilize an $L1_0$ single phase. This intermetallic phase is confirmed to host high magnetocrystalline anisotropy, likely attributed to the lower symmetry of the $L1_0$ structure and the strong spin-orbit coupling provided by the Pt insertion. These findings demonstrate the promise of high-entropy intermetallic design as a key avenue to search for new high-entropy magnets beyond the equiatomic composition.

## 4. Experimental Section

The design of these HEA thin films is based on the estimation of mixing enthalpy $\Delta H_{mix}$ using the Miedema method.[34] A table of binary $\Delta H_{mix}$ for $3d$ metals is given in the Supporting Information.[16] The element-weighted average mixing enthalpy of the solid solution FeCoNiMnCu is $\Delta H_{mix} = 1.23$ kJ mol$^{-1}$. The small positive enthalpy is outweighed by the ideal configurational entropic contribution to the free energy at room-temperature, $-TS_{config} = -3.9$ kJ mol$^{-1}$. The valence electron concentration of 9 for equiatomic FeCoNiMnCu would correspond to a *fcc* structure for $3d$ transition metal HEAs,[18] which is a precursor to the $L1_0$ structure. The system (FeCoNiMnCu)Pt also satisfies the geometric and electronic criteria for single-phase HEIs with atomic size difference $\delta r = 5.7\%$ and modified electronegativity difference $\eta = 0.27$.[12b]

The FeCoNiMnCu films were deposited by direct-current (DC) magnetron sputtering in an ultrahigh vacuum system with a base pressure of ~ 1×10$^{-8}$ Torr. A homemade composite Fe-Co-Ni-Mn-Cu target was used, which was formed by cold-pressing a mixture of elemental powders, with purities of 99.9% (Fe), 99.8% (Co), 99.9% (Ni), 99.6% (Mn), and 99.9% (Cu). The particle size of all powders was under 10 $\mu$m. The powders were mixed in an equimolar ratio, and the mixture was uniaxially pressed with 50 metric tons onto a Cu backing plate, forming a target 2" in diameter and 1/8" thick. Films of 13 nm and 50 nm thickness were sputtered in 2.5 mTorr Ar with 50 W power onto thermally oxidized Si (100) substrates with a 285 nm thick amorphous SiO$_2$ layer. Substrate temperature was held at $T_s$ = 20, 350, 500, 600, and 700 °C. The films were cooled in 30 mTorr of Ar for up to 1 h before deposition of 4-5 nm of Ta or Ti capping layer to prevent oxidation. The (FeCoNiMnCu)Pt films of 20 nm thickness were deposited onto thermally oxidized Si (100) substrates at room temperature in 2 mTorr Ar, by co-sputtering of elemental Fe, Co, Ni, Mn, Cu, and Pt targets using 35 W, 30 W, 28 W, 20 W, 13 W, and 70 W power, respectively.[35] Additional 20 nm (FeCoNiMnCu)Pt films were deposited onto Si/SiO$_2$ by co-sputtering of Pt with the composite FeCoNiMnCu target in 2 mTorr Ar at powers of 20 W and 70 W, respectively. The results are reported in the Supporting Information (**Figures S6** and **S7**).

Sputtered films were treated with RTA in a vacuum chamber with a base pressure ~ 1×10$^{-8}$ Torr. During RTA, the sample was transferred from an adjacent load-lock chamber under vacuum into the main chamber where it is placed directly under a heating lamp for the designated annealing time, ranging from 10 s to 60 s. The annealing temperature of 600 °C corresponds to the equilibrium temperature of the surrounding chamber environment, as set via thermocouple in contact with the lamp window, located approximately 1-2 cm from both the sample and heating filament. Subsequently, the sample was transferred back to the load-lock, which was vented with dry N$_2$ and brought to atmospheric pressure over a period of approximately 2 min, during which the sample is cooled to room-temperature.

X-ray diffraction with Cu K$_\alpha$ radiation was employed for structural characterization using a Panalytical X'Pert$^3$ Materials Research Diffractometer. Surface topography was imaged on Zeiss SUPRA 55-VP scanning electron microscope and NT-MDT atomic force microscope. An Oxford Instruments EDX system was used to analyze chemical composition of the films. High angle annular dark field scanning transmission electron microscopy (HAADF-STEM) and selected area electron diffraction (SAED) were performed on plan-view samples for microstructural analysis.

Magnetic measurements were performed via vibrating sample magnetometry (VSM) using a Princeton Measurements Corporation MicroMag and superconducting quantum interference device (SQUID) magnetometry on a Quantum Design Magnetic Property Measurements System (MPMS3) system. FORC measurements were carried out following prior procedures.[20] The sample was first saturated, then brought to a reversal field ($H_R$), and its magnetization $M(H, H_R)$ measured in an applied field $H$ back to saturation. This process was repeated at successively more negative $H_R$ to create a family of FORCs. A FORC distribution was then extracted using $\rho(H, H_R) \equiv -\frac{1}{2M_S}\frac{\partial^2 M(H,H_R)}{\partial H\, \partial H_R}$, where $M_S$ is the saturation magnetization. The FORC distribution is represented in terms of local coercive field $H_c = \frac{1}{2}(H - H_R)$ and bias field $H_b = \frac{1}{2}(H + H_R)$.[36] The local coercivity often correlates with, but is not necessarily the same as, the major loop coercivity.

First-principles calculations were carried out for bulk HEAs corresponding to the compositions and the lattice constants obtained in experiment. We used density functional theory implemented in Questaal [37] with the Korringa-Kohn-Rostocker (KKR) method [38] and Coherent-Potential Approximation (CPA).[39] Each high-entropy species was assumed to have two spin components that were allowed to have either ferro- or antiferromagnetic order during the self-consistent iterations.

*Statistical Analysis*: Information of the FeCoNiMnCu films studied is provided in Supporting Information, Table S2. Saturation magnetization values determined are represented in the form of mean ± standard deviation. Other experimentally measured values are accurate to the last significant digit.

**Supporting Information**

Supporting Information is available from the Wiley Online Library or from the author.

**Acknowledgements**

This work has been supported by the NSF (ECCS-2151809). The acquisition of the MPMS3 system used in this study was supported by the NSF-MRI program (DMR-1828420). This work used Bridges-2 at Pittsburgh Supercomputing Center through allocation PHY230018 from the Advanced Cyberinfrastructure Coordination Ecosystem: Services & Support (ACCESS) program, which is supported by NSF grants #2138259, #2138286, #2138307, #2137603, and #2138296. H.Z. acknowledges support from the U.S. Department of Commerce, NIST under financial assistance awards 70NANB22H101/MML22-1014. Disclaimer: Certain commercial equipment, instruments, software, or materials are identified in this paper in order to specify the experimental procedure adequately. Such identifications are not intended to imply recommendation or endorsement by NIST, nor it is intended to imply that the materials or equipment identified are necessarily the best available for the purpose.

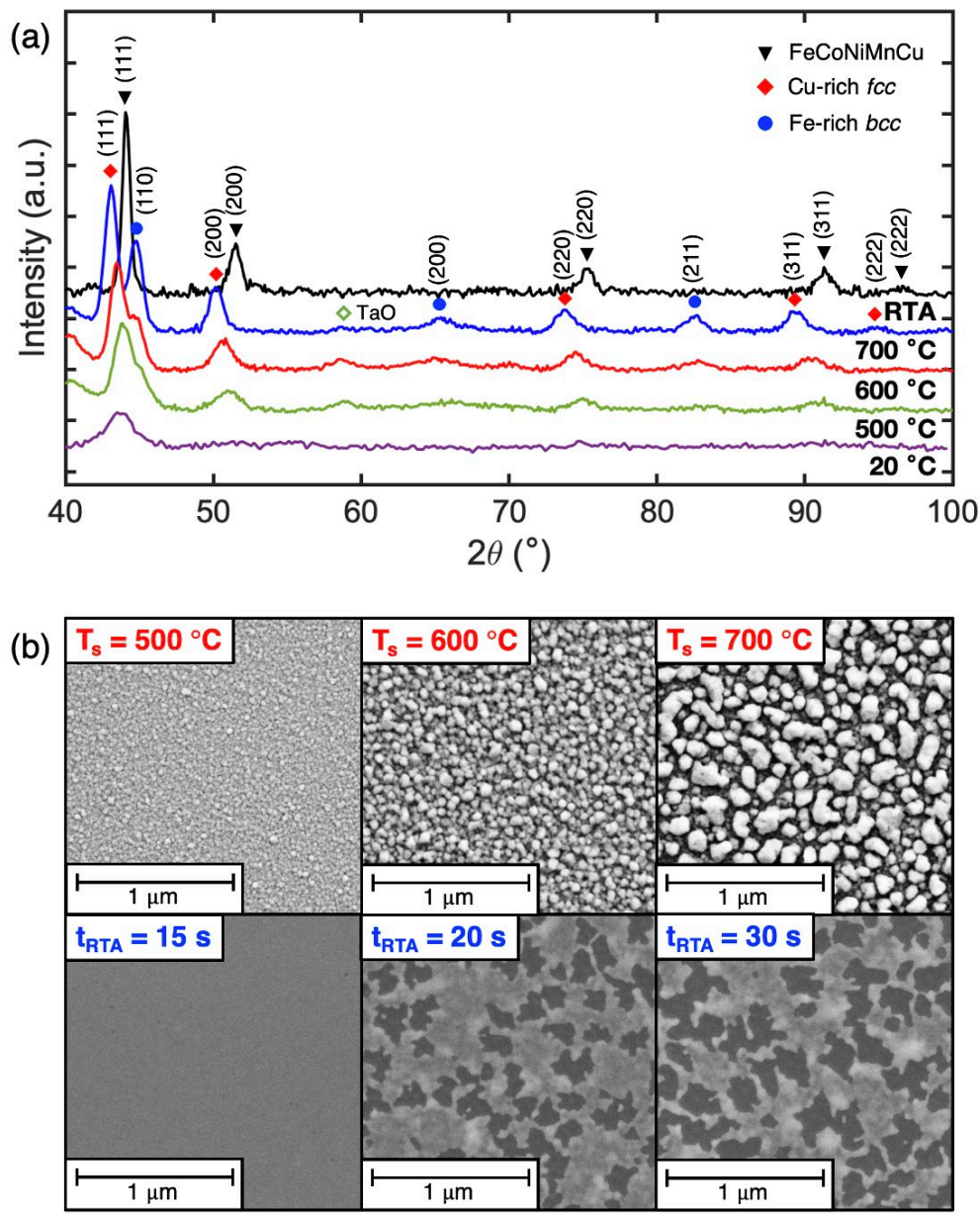

**Figure 1**. **XRD and SEM study of FeCoNiMnCu films grown under different conditions.** (a) GIXRD scan for 50 nm FeCoNiMnCu films sputtered at various substrate temperatures of 20 °C, 500 °C, 600 °C and 700 °C (bottom to top), and sputtered at 20 ° C and annealed by RTA at 600 °C for 30 s (top curve). (b) SEM images for (top) 50 nm films sputtered at substrate temperatures of 500 °C, 600 °C, 700 °C and (bottom) 13 nm films after RTA at 600 °C for 15 s, 20 s, and 30 s.

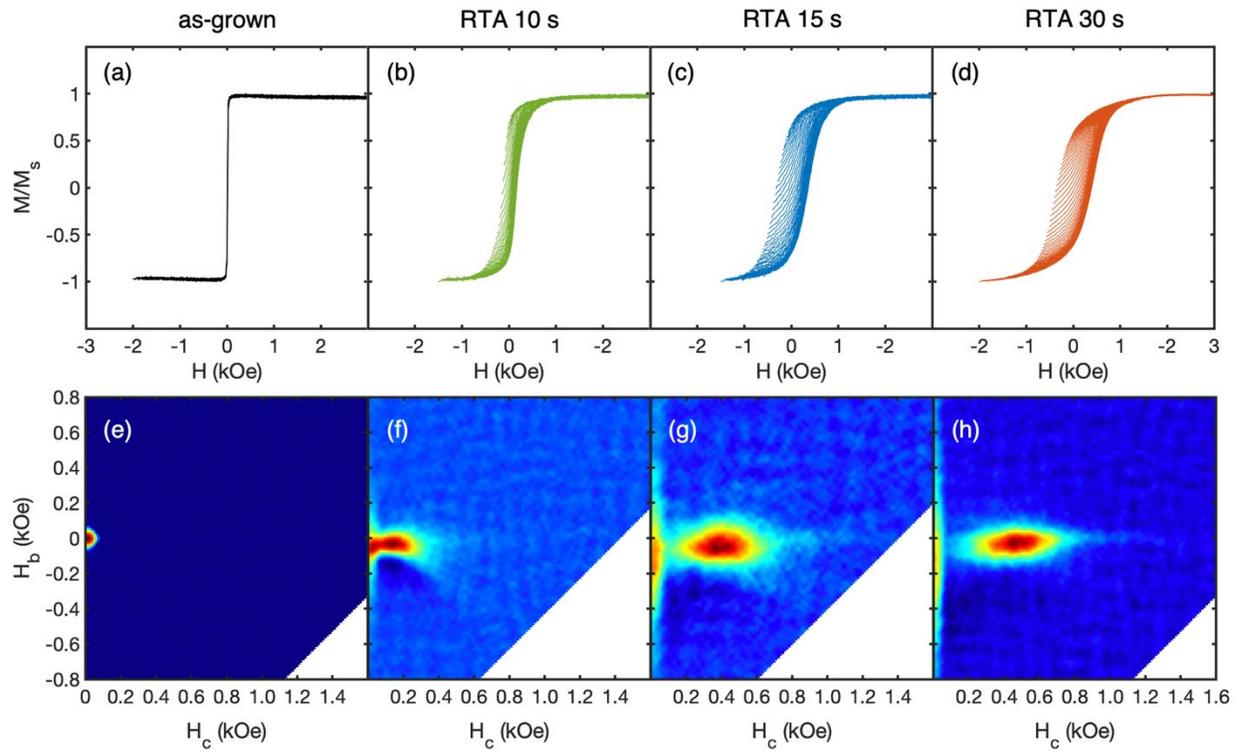

**Figure 2. FORC study of 13 nm FeCoNiMnCu films in the in-plane geometry.** (a-d) Families of FORCs and (e-h) the corresponding FORC distributions (a,e) after sputtering at 350 °C, and after subsequent RTA at 600 °C for (b,f) 10 s, (c,g) 15 s, and (d,h) 30 s.

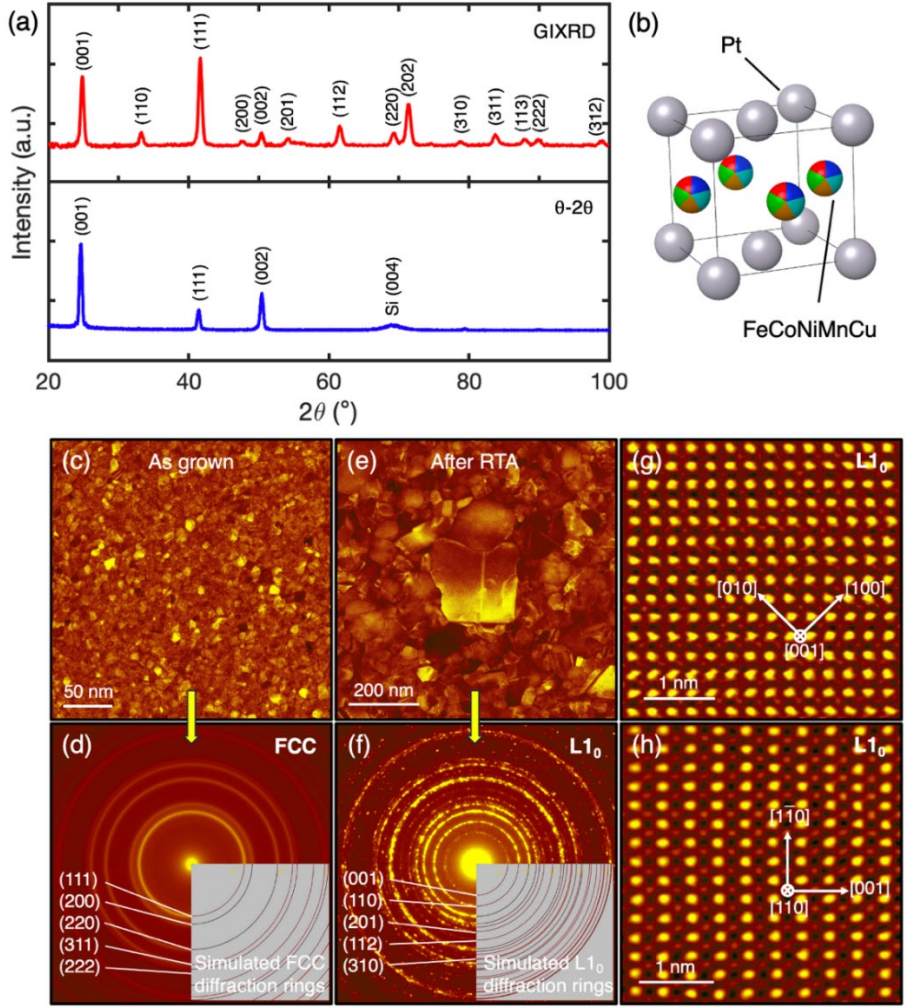

**Figure 3. XRD and plan-view TEM microstructural study of (FeCoNiMnCu)Pt films.** (a) (Top) GIXRD and (bottom) $\theta$-$2\theta$ XRD scans for (FeCoNiMnCu)Pt films after RTA at 600 °C for 60 s. (b) Schematic crystal structure of $L1_0$ (FeCoNiMnCu)Pt. (c) HAADF-STEM image showing uniform nano-grains from as grown sample, and (d) the corresponding SAED pattern showing a well-defined *fcc* structure. (e) HAADF-STEM image showing grown grains from RTA sample, and (f) the corresponding SAED pattern showing a well-defined $L1_0$ structure with indexing of selected forbidden *fcc* reflections. (g, h) Atomic resolution HAADF-STEM images taken along the [001] and [110] zone-axis, respectively, showing the chemical ordering of the $L1_0$ structure on the FeCoNiMnCu and Pt sites.

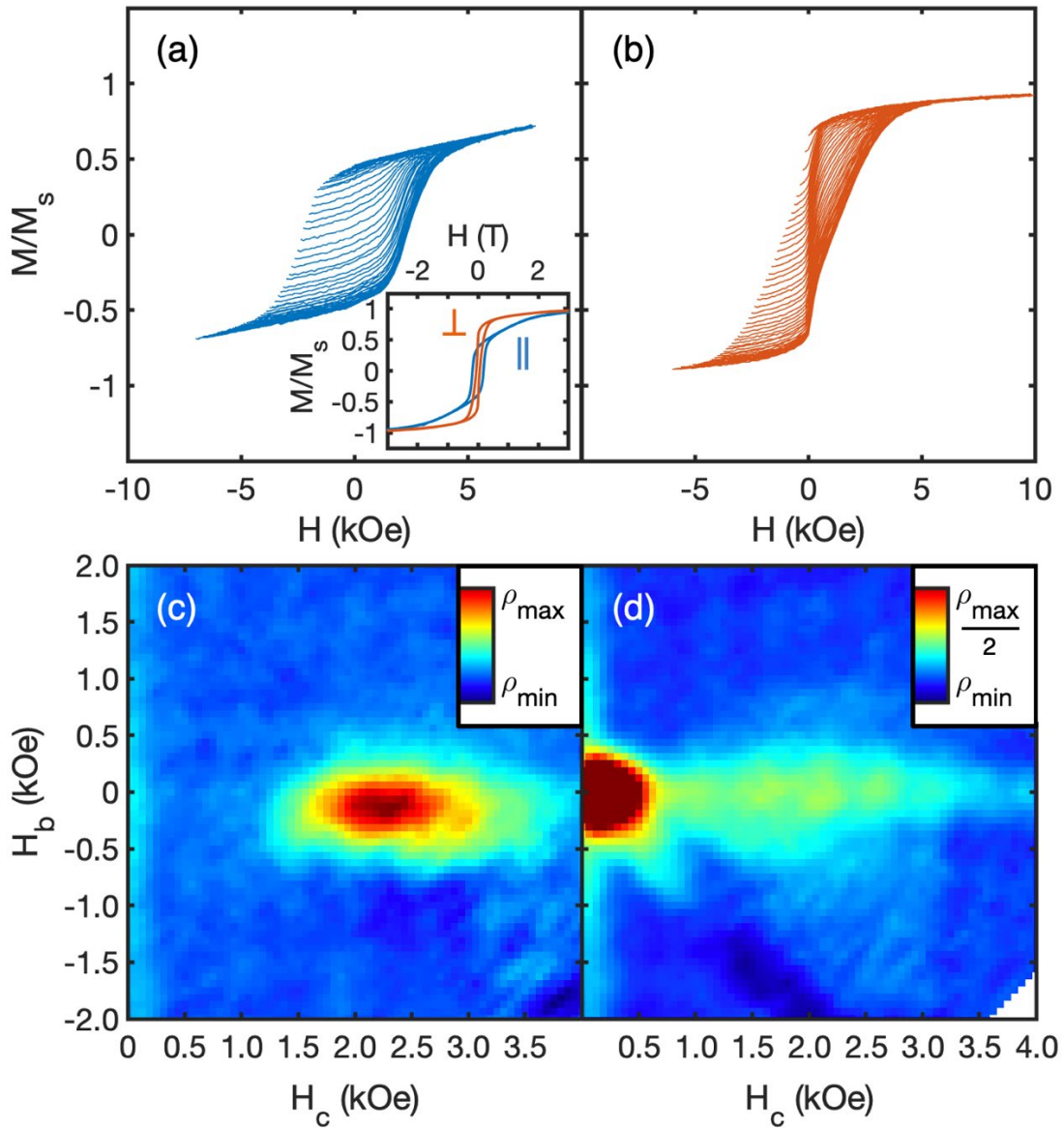

**Figure 4. FORC of (FeCoNiMnCu)Pt film after RTA**. (a) In-plane and (b) out-of-plane family of FORCs and (c,d) corresponding FORC distributions for (FeCoNiMnCu)Pt films after RTA at 600 °C for 60 s. Note the adjustment of color scale in (d). Inset in (a) shows the major loop in the in-plane (blue) and out-of-plane (brown) geometry.

# Supporting Information

**Single-Phase *L*1$_0$-Ordered High Entropy Thin Films with High Magnetic Anisotropy**

Willie B. Beeson,[1] Dinesh Bista,[1] Huairuo Zhang,[2,3] Sergiy Krylyuk,[2] Albert Davydov,[2] Gen Yin,[1] and Kai Liu[1,*]

[1]Department of Physics, Georgetown University, Washington, DC 20057, USA
[2]Materials Science and Engineering Division, National Institute of Standards and Technology, Gaithersburg, MD 20899, USA
[3]Theiss Research, Inc., La Jolla, CA 92037, USA

**Table S1**. Binary $\Delta H_{mix}$ for 3*d* metals.[16] The red lines delineate the Fe-Co-Ni-Mn-Cu system.

|    | Ti      | V       | Cr     | Mn      | Fe      | Co      | Ni      | Cu     | Zn      |
|----|---------|---------|--------|---------|---------|---------|---------|--------|---------|
| Ti | 0       | -1.65   | -7.348 | -8.094  | -16.527 | -27.948 | -34.085 | -8.818 | -22.324 |
| V  | -1.65   | 0       | -1.937 | -0.708  | -7.057  | -13.855 | -17.858 | 4.892  | -8.834  |
| Cr | -7.348  | -1.937  | 0      | 2.141   | -1.448  | -4.44   | -6.645  | 12.385 | -2.247  |
| Mn | -8.094  | -0.708  | 2.141  | 0       | 0.29    | -5.132  | -8.121  | 3.682  | -13.346 |
| Fe | -16.527 | -7.057  | -1.448 | 0.29    | 0       | -0.561  | -1.542  | 12.882 | -3.389  |
| Co | -27.948 | -13.855 | -4.44  | -5.132  | -0.561  | 0       | -0.218  | 6.373  | -12.061 |
| Ni | -34.085 | -17.858 | -6.645 | -8.121  | -1.542  | -0.218  | 0       | 3.495  | -15.802 |
| Cu | -8.818  | 4.892   | 12.385 | 3.682   | 12.882  | 6.373   | 3.495   | 0      | -5.679  |
| Zn | -22.324 | -8.834  | -2.247 | -13.346 | -3.389  | -12.061 | -15.802 | -5.679 | 0       |

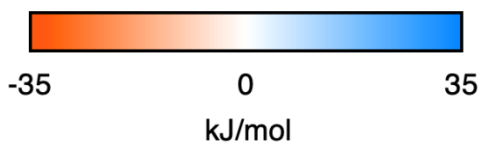

kJ/mol

**Table S2**. Fabrication conditions and magnetic properties of FeCoNiMnCu films.

| Thickness (nm) | Substrate temperature (°C) | RTA time at 600°C (s) | $M_s$ (emu/cm³) | $H_c$ (Oe) |
|---|---|---|---|---|
| 13 | 350 |    | 595 ± 45 | 10 |
|    |     | 10 | 565 ± 45 | 170 |
|    |     | 15 | 640 ± 60 | 350 |
|    |     | 20 | 550 ± 40 | 330⋆ |
|    |     | 30 | 590 ± 30 | 390⋆ |
| 50 | 20  |    | 600 ± 25 | 15 |
|    |     | 30 | 590 ± 20 | 245⋆ |
|    | 500 |    | 665 ± 25 | 80 |
|    | 600 |    | 670 ± 15 | 315⋆ |
|    | 700 |    | 675 ± 25 | 475⋆ |

⋆ denotes films which are discontinuous

Table S2 lists the studied FeCoNiMnCu film fabrication conditions and magnetic properties. A star is added to coercivity values to denote samples that are not continuous films, due to dewetting which produces either voids or isolated islands. The 50 nm films sputtered at elevated temperature exhibit a large coercivity increase, however this is clearly seen to correlate with the morphology change from thin films to isolated nanoislands, as shown in Fig. 1b, which leads to coercivity enhancement via size effects. Meanwhile, the 13 nm films annealed for 10 s and 15 s exhibit comparable coercivity increase while remaining as continuous thin films.

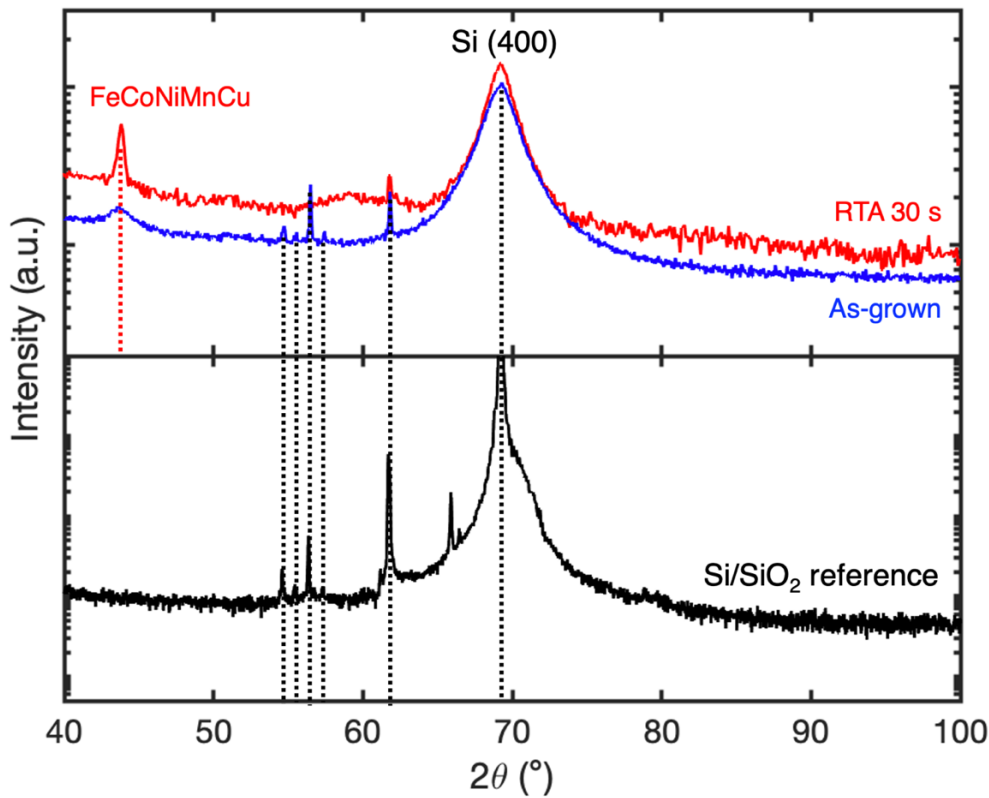

**Figure S2.** (Top) $\theta$-$2\theta$ XRD scan with 1° $\omega$ offset of 50 nm FeCoNiMnCu film grown at 20 °C and after RTA at 600 °C for 30 s. (Bottom) Reference $\theta$-$2\theta$ scan of Si/SiO$_2$ substrate with 0° $\omega$ offset.

### $\theta$-$2\theta$ XRD of FeCoNiMnCu films

$\theta$-$2\theta$ XRD scans with $\omega$ offset of 1° was performed for the 50 nm FeCoNiMnCu films sputtered at 20 °C and treated with RTA for 30 s, as seen in the upper panel of Figure S1. Significant crystal growth after RTA can be seen by the increased intensity and narrow width of the primary peak at 44°. The film after RTA appears to show a preferred (111) texture, due to the lack of significant *fcc* peaks at higher $2\theta$. The lower panel of Figure S1 shows a reference $\theta$-$2\theta$ scan of the Si/SiO$_2$ substrate, accounting for the peaks at 55° and above. Note that no $\omega$ offset is used in the reference scan, leading to sharp Si (400) diffraction peaks for Cu K$_\alpha$, Cu K$_\beta$, and W L$_\alpha$ wavelengths which are absent in the upper panel.

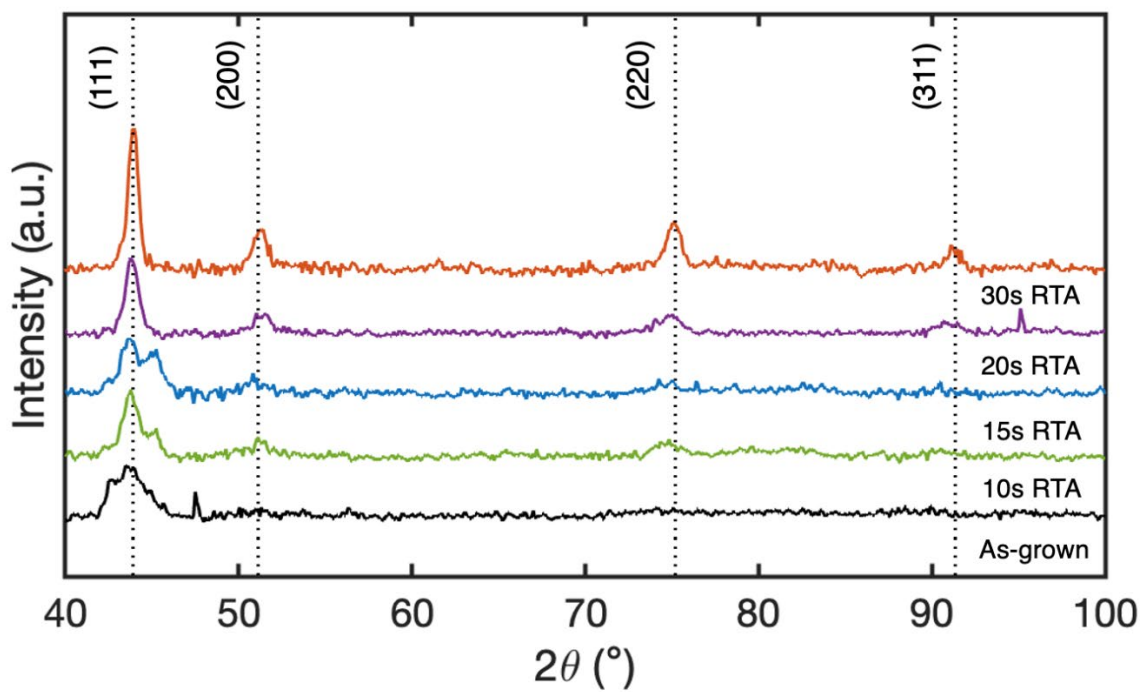

**Figure S3.** GIXRD scans ($\omega = 0.4°$) for 13 nm FeCoNiMnCu films sputtered at 350 °C and after RTA treatment at 600 °C for 10 s, 15 s, 20 s, and 30 s.

**GIXRD of 13 nm FeCoNiMnCu for different annealing times**

Figure S2 shows the GIXRD scans ($\omega = 0.4°$) for the 13 nm FeCoNiMnCu films grown at 350 °C and treated with RTA for different times. The films treated with RTA were capped with 4 nm Ti layer. The as-grown film shows a broad primary peak at $2\theta = 44°$, similar to the films grown at 20 °C. The films treated with RTA for 10 s and 15 s exhibit *fcc* peaks in addition to a peak at ~45°, which is consistent in position with the (110) peak of the *bcc* phase observed in the 50 nm films grown at high temperature (Figure 1). The films annealed for 20 s and 30 s show only *fcc* phase peaks.

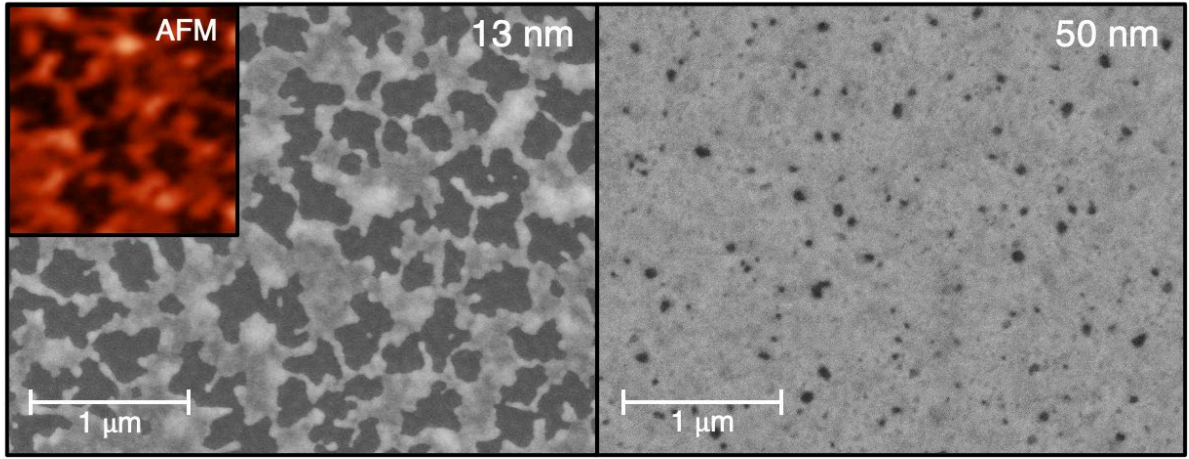

**Figure S3.** SEM images of FeCoNiMnCu films after 30 s of RTA for thicknesses of 13 nm (left) and 50 nm (right). The inset on the left is the AFM measurement of the film surface, showing the height difference corresponding to SEM contrast. The scale bar of the AFM image is the same as the SEM.

**Effect of film thickness on void growth**

SEM images taken after 30 s RTA time reveal the difference in void growth between 13 nm and 50 nm film. The inset for the SEM image of 13 nm film (left) shows the AFM measurement of the sample surface, where the height variation can be seen to match the contrast in SEM. ImageJ software was used to analyze the density and area fraction of voids. In the 13 nm film, voids make up 45% area fraction and have a number density of approximately 5 $\mu m^{-2}$, while the voids in 50 nm film make up less than 5% area fraction but with a comparable number density of approximately 8 $\mu m^{-2}$. The difference in number density could be attributed to the merging of voids upon growth. The comparable number density but smaller area fraction reflects the vastly suppressed void growth kinetics in the thicker film, while the void nucleation rate is unchanged.

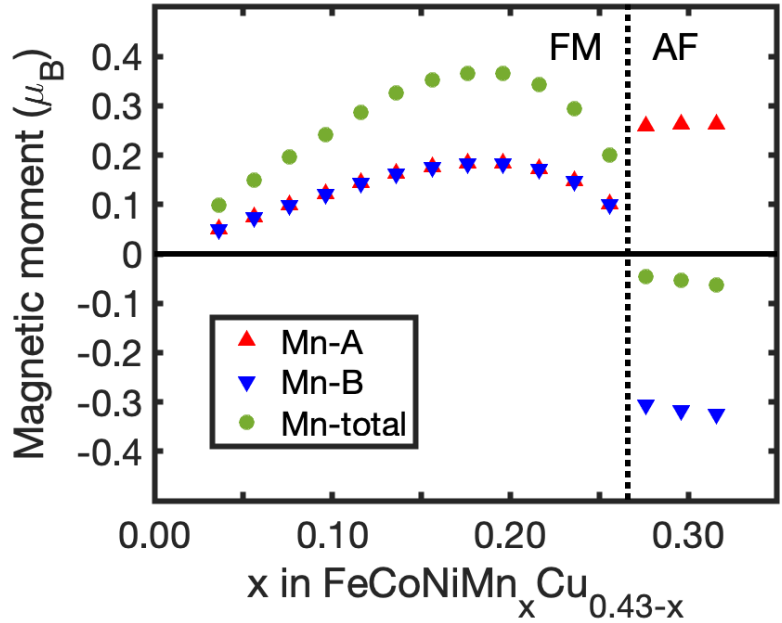

**Figure S4.** Calculated magnetic moments of Mn in $Fe_{0.19}Co_{0.18}Ni_{0.20}Mn_xCu_{0.43-x}$ given by DFT. The magnetic configuration of the random Mn sites is explored by allowing 50% of Mn sites (Mn-A, red up triangle) to host either FM or AF moments compared to the other 50% (Mn-B, blue down triangle). Mn-total (green circle) represents the sum of moments for Mn-A and Mn-B at the ground state.

**Calculated magnetic moment of Mn in $Fe_{0.19}Co_{0.18}Ni_{0.20}Mn_xCu_{0.43-x}$.**

We have used density functional theory (DFT) to understand the ground-state magnetization configuration for the near equiatomic $Fe_{0.19}Co_{0.18}Ni_{0.20}Mn_xCu_{0.43-x}$. We have employed the DFT implemented in Questaal using the Korringa-Kohn-Rostocker (KKR) method and the mean-field Coherent-Potential Approximation (CPA). In this calculation we allow each high-entropy element to have half of random sites being AF or FM aligned with the other half, and search for a lower ground-state energy. As shown in Figure S4, a FM to AF transition for Mn sites is found when changing the concentration ratio of Mn:Cu (Figure S4). All other elements, however, are found to have pure FM phase regardless of the Mn concentration.

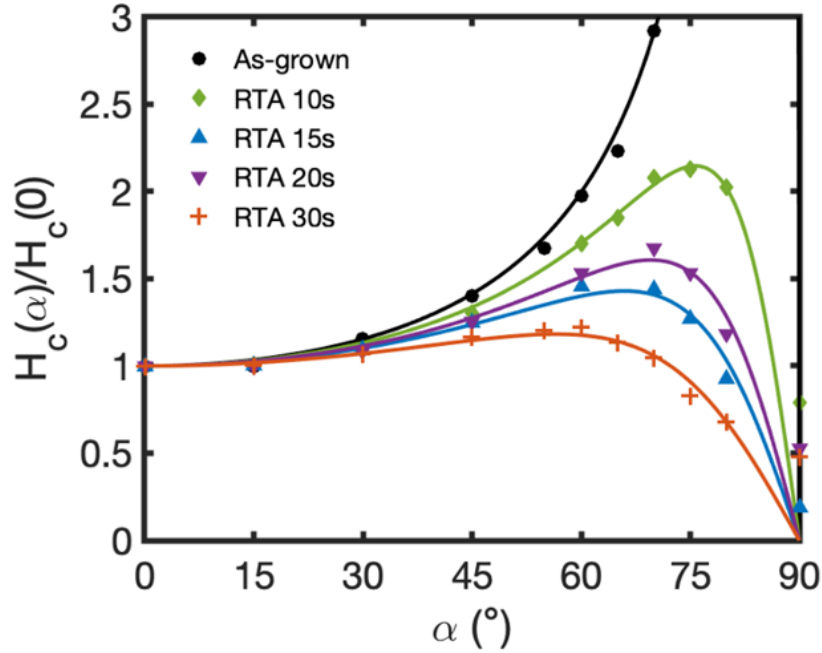

**Figure S5.** Dependence of coercivity on the angle $\alpha$ between $H$ and the film plane for the FeCoNiMnCu films after RTA for different times. The fits of the data to Equation 1 are shown by the solid lines.

**Angular dependence of coercivity**

The angular variation in coercivity was measured to gain further insight into the evolution of magnetic properties. Figure S5 shows the dependence of the normalized coercivity $H_c(\alpha)/H_c(0)$ on the angle $\alpha$ between the field direction and film surface. $H_c(0)$ is the coercivity measured in-plane. The angular variation in coercivity for multi-domain crystals due to the combination of domain wall motion and coherent reversal may be described by a generalized Kondorsky model:[28]

$$\frac{H_c(\alpha)}{H_c(0)} = \frac{\cos\alpha}{N\sin^2\alpha + \cos^2\alpha} \quad (1)$$

where $N$ is a parameter relating the demagnetization factors as $N = \frac{N_x}{N_z + N_A}$, with $N_z$ and $N_x = N_y$ being the demagnetization factors in the out-of-plane ($\alpha = 90°$) and in-plane ($\alpha = 0°$) directions, respectively, and more generally the $\alpha = 0°$ defines the easy-axis; $N_A$ represents an effective demagnetization factor due to uniaxial anisotropies other than the shape. The equation reduces to the Kondorsky relation $(\cos\alpha)^{-1}$ for $N_x = 0$. The model predicts $H_c(90°) = 0$ for perfect single-crystals, whereas $H_c(90°) \neq 0$ for polycrystals. Excluding $H_c(90°)$, the data is reasonably well-fit to Equation 1, which indicates the pinning-controlled coercivity mechanism for angles close to in-plane. The deviation from $H_c \propto (\cos \alpha)^{-1}$ at higher angles signals a transition to coherent rotation or nucleation-controlled coercivity. In the as-grown state, $N = 0.0007$, reflecting the extreme thin film shape demagnetization factors, while RTA treatment for 15 s leads to an increase of $N$ to 0.143. After 20 s of RTA, there is a slight decrease to 0.108 then finally an increase to 0.233 after 30 s of RTA. Here, an increase in $N$ effectively indicates the reduction of in-plane anisotropy. This could relate to shape anisotropy changes as a result of dewetting in the case of the films annealed for 20 s and 30 s. However, a substantial change of $N$ is observed after 15 s RTA without any obvious change in film morphology. In this case, the increase of $N$ could reflect local changes in shape anisotropy or other forms of strain-induced anisotropy leading to a negative value of $N_A$. Notably, upon film dewetting at 20 s RTA, there is a slight decrease of $N$, while the morphology changes alone would be expected to increase $N$. Considering the observed reduction of residual stress upon film dewetting in similar RTA-treated films,[20a] this behavior is consistent with the proposed role of strain on the modification of anisotropy in the present films. The overall trend observed here illuminates the formation of the reversible ridge in the in-plane FORC distributions due to an increase of demagnetization effects in the film plane.

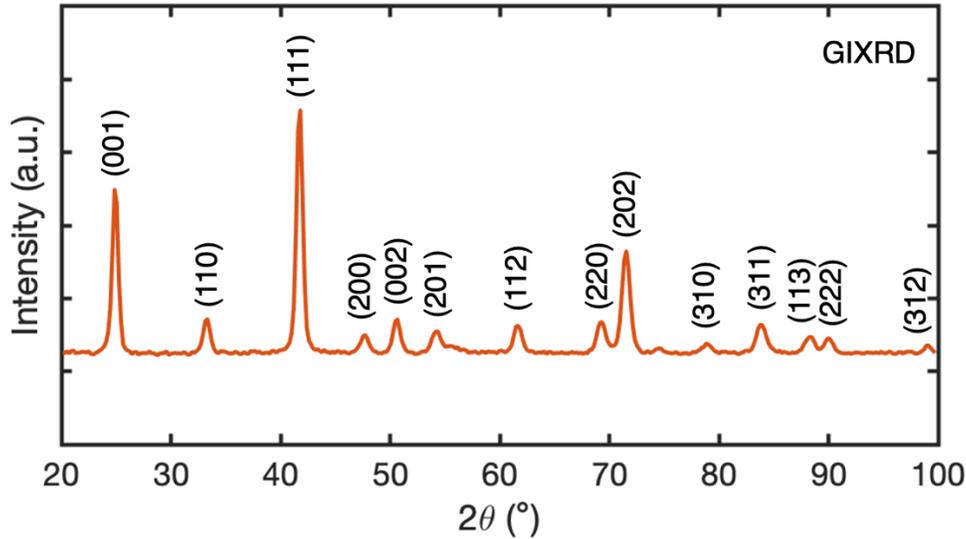

**Figure S6.** GIXRD scans ($\omega = 0.4°$) of 20 nm (FeCoNiMnCu)Pt films deposited by co-sputtering of the composite FeCoNiMnCu and elemental Pt targets and treated with RTA at 600 °C for 60 s.

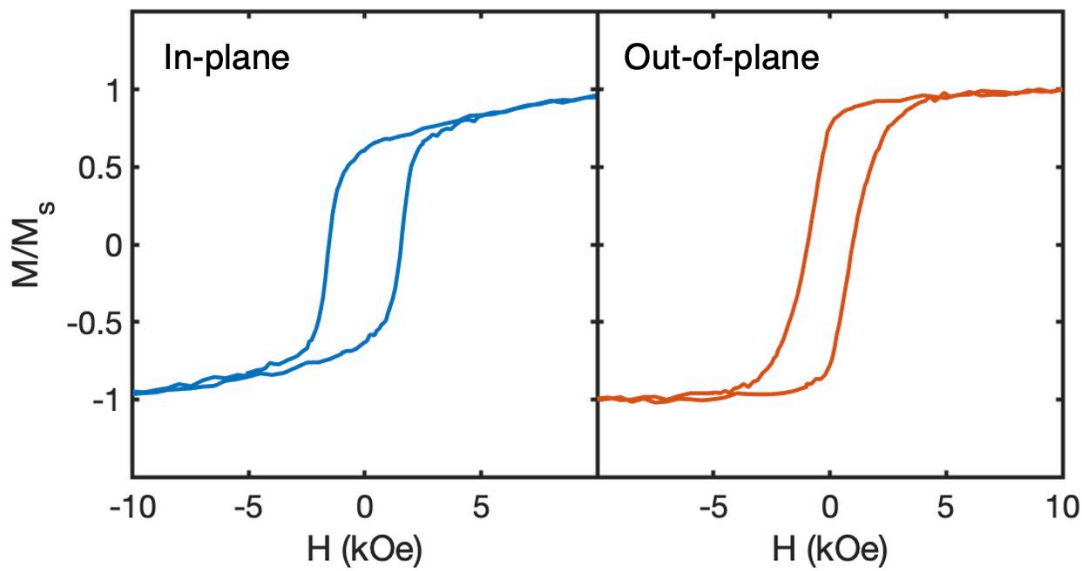

**Figure S7.** Hysteresis loops of the 20 nm (FeCoNiMnCu)Pt film sputtered using the composite target after RTA at 600 °C for 60 s for the in-plane and out-of-plane geometry.

**(FeCoNiMnCu)Pt films deposited by co-sputtering of Pt with a composite FeCoNiMnCu target**

Additional (FeCoNiMnCu)Pt films were deposited by co-sputtering of Pt with the composite target formed by powder compaction. EDX analysis of the composite target film gives an average composition of $Fe_{0.11}Co_{0.09}Ni_{0.15}Mn_{0.07}Cu_{0.12}Pt_{0.46}$, vs. $Fe_{0.11}Co_{0.12}Ni_{0.10}Mn_{0.09}Cu_{0.10}Pt_{0.48}$ for the film deposited by co-sputtering of elemental targets. Figure S6 shows the GIXRD scans ($\omega = 0.4°$) for the 20 nm (FeCoNiMnCu)Pt films deposited by composite target method after RTA treatment for 60 s at 600 °C. The XRD pattern is very similar to that of Fig. 6a, indicating that the $L1_0$ structure is readily obtained by post-annealing using either method. The composite target film has slightly smaller lattice parameters of $a = 3.57$ Å and $c = 3.75$ Å compared with the film sputtered using the elemental targets ($a = 3.60$ Å and $c = 3.83$ Å), which may be due to the slight differences in composition, particularly the lower Pt content. Figure S7 shows the major hysteresis loops of the composite target sample measured using VSM for the in-plane and out-of-plane geometry, with coercivities of 1.60 kOe and 0.96 kOe, respectively. These coercivities are of comparable magnitude but somewhat smaller than the sample sputtered using only elemental targets. This decrease may be due to the difference in composition, specifically the higher Ni content, as addition of Ni in FePt is known to steadily reduce the anisotropy due to the increase in effective 3$d$ electron density.[32, 40] Here, the effective electron density is increased from 9.4 in the elemental target film to 10 in the composite target film. In summary, the (FeCoNiMnCu)Pt films deposited by co-sputtering of Pt with either elemental targets or a composite target both exhibit similar structure and magnetic properties after RTA, with small differences attributed to changes in film composition. Deposition using a composite target has the advantage of requiring fewer sputtering guns. However, deposition by co-sputtering of elemental targets allows convenient and fine control over the film composition by adjusting sputtering power on each source.